\documentclass[journal]{IEEEtran}
\def\baselinestretch{0.81}

\ifCLASSINFOpdf
\else
\fi
\usepackage{amsmath}
\usepackage{amssymb,xspace,intmacros,subfigure,amsfonts}
\usepackage[final]{graphicx}
\usepackage{color}
\newcommand{\xb}{{\textbf{x}}}

\newcommand{\Fb}{{\textbf{F}}}

\newcommand{\cb}{{\textbf{c}}}
\newcommand{\Cb}{{\textbf{C}}}

\newcommand{\bb}{{\textbf{b}}}

\newcommand{\Ib}{{\textbf{I}}}

\newcommand{\hb}{{\textbf{h}}}
\newcommand{\Hb}{{\textbf{H}}}

\begin{document}
%
% paper title
% can use linebreaks \\ within to get better formatting as desired
% Do not put math or special symbols in the title.
\title{Double-detector for Sparse Signal Detection from One Bit Compressed Sensing Measurements}
%
%
% author names and IEEE memberships
% note positions of commas and nonbreaking spaces ( ~ ) LaTeX will not break
% a structure at a ~ so this keeps an author's name from being broken across
% two lines.
% use \thanks{} to gain access to the first footnote area
% a separate \thanks must be used for each paragraph as LaTeX2e's \thanks
% was not built to handle multiple paragraphs
%

\author{Hadi~Zayyani,
Farzan~Haddadi,~\IEEEmembership{Member,~IEEE,}~and Mehdi~Korki,~\IEEEmembership{Student Member,~IEEE}
        %and~Farrokh~Marvasti,~\IEEEmembership{Senior Member,~IEEE}% <-this % stops a space
\thanks{H. Zayyani is with the Department
of Electrical and Computer Engineering, Qom University of Technology, Qom, Iran (e-mail: zayyani@qut.ac.ir).}% <-this %
\thanks{F. Haddadi is with the Department
of Electrical Engineering, Iran University of Science and Technology, Tehran, Iran (e-mail: fhaddadi@iust.ac.ir).}
\thanks{M. Korki is with School of Software and Electrical Engineering, Swinburne University of Technology, Hawthorn,
3122 Australia (e-mail: mkorki@swin.edu.au).}

%stops a space
%\thanks{J. Doe and J. Doe are with Anonymous University.}% <-this % stops a space
%\thanks{Manuscript received April 19, 2005; revised December 27, 2012.}
}

% note the % following the last \IEEEmembership and also \thanks -
% these prevent an unwanted space from occurring between the last author name
% and the end of the author line. i.e., if you had this:
%
% \author{....lastname \thanks{...} \thanks{...} }
%                     ^------------^------------^----Do not want these spaces!
%
% a space would be appended to the last name and could cause every name on that
% line to be shifted left slightly. This is one of those "LaTeX things". For
% instance, "\textbf{A} \textbf{B}" will typeset as "A B" not "AB". To get
% "AB" then you have to do: "\textbf{A}\textbf{B}"
% \thanks is no different in this regard, so shield the last } of each \thanks
% that ends a line with a % and do not let a space in before the next \thanks.
% Spaces after \IEEEmembership other than the last one are OK (and needed) as
% you are supposed to have spaces between the names. For what it is worth,
% this is a minor point as most people would not even notice if the said evil
% space somehow managed to creep in.

% The paper headers
\markboth{IEEE Signal Processing Letters,~Vol.XX, No.X}%
{Shell \MakeLowercase{\textit{et al.}}:}
% The only time the second header will appear is for the odd numbered pages
% after the title page when using the twoside option.
%
% *** Note that you probably will NOT want to include the author's ***
% *** name in the headers of peer review papers.                   ***
% You can use \ifCLASSOPTIONpeerreview for conditional compilation here if
% you desire.

% If you want to put a publisher's ID mark on the page you can do it like
% this:
%\IEEEpubid{0000--0000/00\$00.00~\copyright~2012 IEEE}
% Remember, if you use this you must call \IEEEpubidadjcol in the second
% column for its text to clear the IEEEpubid mark.

% use for special paper notices
%\IEEEspecialpapernotice{(Invited Paper)}

% make the title area
\maketitle

% As a general rule, do not put math, special symbols or citations
% in the abstract or keywords.
\begin{abstract}
This letter presents the sparse vector signal detection from one bit compressed sensing measurements, in contrast to the previous works which deal with scalar signal detection. In this letter, available results are extended to the vector case and the GLRT detector and the optimal quantizer design are obtained. Also, a double-detector scheme is introduced in which a sensor level threshold detector is integrated into network level GLRT to improve the performance. The detection criteria of oracle and clairvoyant detectors are also derived. Simulation results show that with careful design of the threshold detector, the overall detection performance of double-detector scheme would be better than the sign-GLRT proposed in \cite{Fang13} and close to oracle and clairvoyant detectors. Also, the proposed detector is applied to spectrum sensing and the results are near the well known energy detector which uses the real valued data while the proposed detector only uses the sign of the data.

\end{abstract}

% Note that keywords are not normally used for peerreview papers.

% For peer review papers, you can put extra information on the cover
% page as needed:
 \ifCLASSOPTIONpeerreview
 \begin{center} \bfseries EDICS: MLSAS-SPARSE \end{center}
 \fi
%
% For peerreview papers, this IEEEtran command inserts a page break and
% creates the second title. It will be ignored for other modes.
\IEEEpeerreviewmaketitle

\section{Introduction}
% The very first letter is a 2 line initial drop letter followed
% by the rest of the first word in caps.
%
% form to use if the first word consists of a single letter:
% \IEEEPARstart{A}{demo} file is ....
%
% form to use if you need the single drop letter followed by
% normal text (unknown if ever used by IEEE):
% \IEEEPARstart{A}{}demo file is ....
%
% Some journals put the first two words in caps:
% \IEEEPARstart{T}{his demo} file is ....
%
% Here we have the typical use of a "T" for an initial drop letter
% and "HIS" in caps to complete the first word.

\IEEEPARstart{W}{e} study the problem of decentralized detection in a wireless sensor network (WSN). In WSNs, a set of nodes is required to decide between two hypotheses based on a reduced form of measurements sent to the fusion center (FC) for a final decision. Decentralized detection has experienced a flourishing interest among signal processing research community \cite{Fang13}-\cite{Niu08}. In this case, a local likelihood ratio test (LRT) can be conducted in each sensor and the local decision is sent to a FC to reach a global decision \cite{Fang13}.

In this letter, we study the problem of detecting the presence of an unknown signal from linear measurements. A common practice in decentralized detection is to send the sensor's original observations to the FC. Then, a Generalized Likelihood Ratio Test (GLRT) is used to make a final decision. However, communication between nodes and FC are very bandwidth and energy demanding. Therefore, to save the resources of the sensor network, each sensor quantizes its measurements into one bit of information. Then, a GLRT detector based on one bit information is conducted at FC to obtain the final decision. Multisensor GLRT fusion based on one-bit quantized data is also studied in \cite{Fang13}, \cite{Ciu13}. In \cite{Niu08}, a simple and non optimal fusion rule is suggested to detect an unknown signal with quantized data. In \cite{Fang13}, a GLRT detector is proposed to detect a scalar signal from one bit measurements while a one-bit quantizer design is also provided. A Rao test as an asymptotic surrogate for the GLRT is proposed for multisensor fusion in \cite{Ciu13}.

In this letter, we consider the problem of detecting a sparse signal vector from one bit information of linear measurements. This problem is particularly related to one bit compressed sensing (CS) as an extreme case of quantized CS \cite{ZymnBC10}-\cite{Li15}. Classical CS \cite{CandT06}, \cite{Dono06} neglects the quantization process assuming that the measurements are real values. However, in practice the measurements should be quantized to some discrete levels. This is known as quantized CS \cite{ZymnBC10}. In the extreme case of one bit CS, there are only two discrete levels. Application of one bit CS in WSNs is investigated in \cite{Chen15}.

%One bit compressed sensing has gained much attention in the research community recently \cite{BoufB08}-\cite{Li15}, specially in wireless sensor networks \cite{Chen15}. In the one bit compressed sensing framework, an accurate and stable recovery can be achieved by using only the sign of linear measurements \cite{JacqLBB13}.

In this letter, detection of a sparse vector signal is addressed rather than the scaler case studied in \cite{Fang13}. This model, without the sign measurements, is also used in the context of distributed detection and estimation \cite{CatS10}, \cite{CatS11}. The proposed sparse vector model extends the applicability of the conventional scalar model. For instance, this new model can be applied to spectrum sensing as shown in Section VI. We generalize and simplify the GLRT detector of \cite{Fang13}, based on the new model and present the Cramer-Rao bound for this generalized GLRT detector. We refer to the proposed sparse vector model as sign-GLRT, whose scalar model proposed in \cite{Fang13}. The quantizer design is provided and the optimum quantizer thresholds are shown to be zero. In this generalized problem, one could use the solution of efficient one bit CS algorithms such as Binary Iterative Hard Thresholding (BIHT) \cite{JacqLBB13}, instead of Maximum Likelihood (ML) parameter estimates in GLRT detector. This is because, here, we deal with sparse vector and ML estimate does not consider sparsity constraints while sparsity-driven algorithms, e.g. BIHT do, and thus the parameter estimation performance will be more accurate. Moreover, a double-detector algorithm is suggested which uses an internal threshold detector for the data of each sensor. Then, a final GLRT detector combines the decisions of the sensors. Clairvoyant and oracle detectors are provided with which the performance of double-detector and sign-GLRT detectors are compared. Simulation results show that with carefully designing the internal threshold detectors, the double-detector outperforms the sign-GLRT detector. Moreover, application of the proposed detectors in spectrum sensing, shows close results to the full-data energy detector while it needs very simple structure for WSN.
%The rest of the letter is organized as follows. Section~\ref{sec: Problem} introduces the problem formulation. The GLRT detector and the optimal quantizer design is discussed in Section~\ref{sec: GLRT}. A double GLRT detector is suggested in Section~\ref{sec: doubleGLRT}. Section~\ref{sec: clair} introduces some clairvoyant and oracle detectors. Simulation results are presented in Section~\ref{sec: Sim}. Finally, conclusions are drawn in Section~\ref{sec: con}.
\section{The problem formulation}
\label{sec: Problem}
Consider a sparse vector signal $\boldsymbol{\theta}=[\theta_1,\theta_2,...,\theta_M]^T$ which only a small fraction of its elements are nonzero. Also, consider a set of $N$ sensors in the WSN. Each sensor observes a linear measurement of the sparse vector. The problem is a binary hypothesis testing to detect the possible presence of the unknown sparse signal $\boldsymbol{\theta}$. The binary hypothesis test is to decide between the following two hypotheses:
\begin{align*}
\mathrm{H_0}:\quad x_n&=\omega_n,\\
\mathrm{H_1}:\quad x_n&=\hb^T_n\boldsymbol{\theta}+\omega_n,\quad n=1,2,...,N,
\end{align*}
where $x_n$ is the $n$'th sensor's measurement, $\hb_n\in\mathbb{R}^{M\times1}$ is the known measurement vector at sensor $n$ and $\omega_n$ denotes the additive white Gaussian noise with zero mean and variance $\sigma^2_n$ at sensor $n$. The noise is assumed to be independent across sensors. To meet the stringent power and bandwidth constraints in WSN, each real-valued sensor measurement $x_n$ is quantized into one bit of information. With quantization thresholds $\tau_n$, the binary data $b_n$ is given by
\begin{equation}
b_n=\mathrm{sgn}(x_n-\tau_n),
\end{equation}
where $\mathrm{sgn}(x)=1$ if $x>0$, otherwise $\mathrm{sgn}(x)=0$. Then, the binary data is transmitted to the FC with an ideal binary channel. Therefore, upon receiving the binary data $\{b_n\}_{n=1}^N$, the FC decides between the presence or absence of the sparse vector $\boldsymbol{\theta}$. Through steps similar to \cite{Fang13}, we aim to determine the optimal quantization thresholds and to develop a detector for the sparse signal vector.

\section{GLRT detector and optimal Quantizer design}
\label{sec: GLRT}
Suppose that the quantization thresholds are known for each sensor. Then, a GLRT \cite{Kay98} can be used to detect the presence of the sparse vector $\boldsymbol{\theta}$. The GLRT detection criterion is
\begin{equation}
\label{eq: GLRTbasic}
\mathrm{T}_{\mathrm{Q}}(\bb)\triangleq\frac{P(\bb|\hat{\boldsymbol{\theta}},\mathrm{H}_1)}{P(\bb|\mathrm{H}_0)}\substack{H_1 \\ > \\ < \\ H_0}\eta,
\end{equation}
where $\bb=[b_1,b_2,...,b_N]^T$ is the one bit measurement vector, $\hat{\boldsymbol{\theta}}$ is the ML Estimator (MLE) of $\boldsymbol{\theta}$, the subscript '$\mathrm{Q}$' stands for one bit quantization scheme, $P(\bb|\hat{\boldsymbol{\theta}},\mathrm{H}_1)$ and $P(\bb|\mathrm{H}_0)$ denote the conditional probability distribution function under hypotheses $\mathrm{H}_1$ and $\mathrm{H}_0$, respectively, and $\eta$ is a threshold determined by the false alarm probability. The MLE of $\boldsymbol{\theta}$ is computed by maximizing the log-likelihood ratio function which is $\hat{\boldsymbol{\theta}}=\text{arg }\underset{\boldsymbol{\theta}}{\text{max}} L(\boldsymbol{\theta})$,
%\begin{equation}
%\hat{\boldsymbol{\theta}}=\text{arg }\underset{\boldsymbol{\theta}}{\text{max}} L(\boldsymbol{\theta})
%\end{equation}
where the log-likelihood function can be written as \cite{Fang13}:
\begin{equation*}
L(\boldsymbol{\theta})\triangleq \mathrm{log} P(\bb;\boldsymbol{\theta})=\sum_{n=1}^N\{b_n \mathrm{log}[F_{\omega_n}(\tau_n-\hb^T_n\boldsymbol{\theta})]
\end{equation*}
\begin{equation}
+(1-b_n)\mathrm{log}[1-F_{\omega_n}(\tau_n-\hb^T_n\boldsymbol{\theta})]\}
\end{equation}
and $\{b_n\}$ are independent with probability mass function (PMF) of:
\begin{equation}
\label{eq: pbn}
P(b_n;\boldsymbol{\theta})=[F_{\omega_n}(\tau_n-\hb^T_n\boldsymbol{\theta})]^{b_n}[1-F_{\omega_n}(\tau_n-\hb^T_n\boldsymbol{\theta})]^{1-b_n},
\end{equation}
where $F_{\omega_n}$ denotes the Complementary Cumulative Density Function (CCDF) of $\omega_n$. The MLE of $\boldsymbol{\theta}$ does not have a closed form even in the scalar case discussed in \cite{Fang13}. Yet, an optimization-based approach to find $\hat{\boldsymbol \theta}_{ML}$ is feasible in case the log likelihood function is concave. Since $\omega_n$ follows a Gaussian distribution $F_{\omega_n}$ is log-concave and thus the log-likelihood function in (3) is concave \cite{Fang13}, \cite{ZymnBC10}.
However, the MLE of $\boldsymbol{\theta}$ does not consider the sparsity constraint. Hence, using a sparsity-driven approach to estimate $\boldsymbol{\theta}$ leads to more accurate result.
In this letter, $\boldsymbol{\theta}$ is assumed to be sparse, and thus we can estimate it, more accurately than MLE, by one bit compressed sensing algorithms. One of the simplest, yet efficient algorithms is BIHT \cite{JacqLBB13}. We use the result of the BIHT algorithm instead of the MLE in the GLRT detector.

To design the optimal quantizer and find the optimal value of $\tau_n$, we analyze the asymptotic performance of the GLRT detector. The modified test statistic $2\mathrm{ln}T_{Q}(\bb)$ exhibits the following distribution
\begin{equation}
2\mathrm{ln}T_{Q}(\bb)\sim \Big\{\begin{array}{cc}
                                     \chi^2_1\quad\quad\quad \text{under} H_0 &  \\
                                     \chi^{'2}_1(\lambda_{Q})\quad\text{under} H_1 &
                                    \end{array},
\end{equation}
where $\chi^2_v$ represents a central chi-squared distribution with $v$ degrees of freedom, and $ \chi^{'2}_v(\lambda)$ denotes the non-central chi-squared distribution with $v$ degrees of freedom and noncentrality parameter $\lambda$. The noncentrality parameter $\lambda_{Q}$ is computed as \cite{Kay98} $\lambda_{Q}=(\boldsymbol{\theta}_1-\boldsymbol{\theta}_0)^T\Ib(\boldsymbol{\theta}_0)(\boldsymbol{\theta}_1-\boldsymbol{\theta}_0)$,
%\begin{equation}
%\lambda_{Q}=(\boldsymbol{\theta}_1-\boldsymbol{\theta}_0)^T\Ib(\boldsymbol{\theta}_0)(\boldsymbol{\theta}_1-\boldsymbol{\theta}_0)
%\end{equation}
where $\boldsymbol{\theta}_0=\boldsymbol{0}$ and $\boldsymbol{\theta}_1=\boldsymbol{\theta}$ are the values of $\boldsymbol{\theta}$ under hypotheses $H_0$ and $H_1$, respectively, and $\Ib(\boldsymbol{\theta})$ is the Fisher Information Matrix (FIM) given by \cite{Fang13} $\Ib(\boldsymbol{\theta})=\sum_{n=1}^N\frac{p^2_{\omega_n}(\tau_n-\hb^T_n\boldsymbol{\theta})}{F_{\omega_n}(\tau_n-\hb^T_n\boldsymbol{\theta})(1-F_{\omega_n}(\tau_n-\hb^T_n\boldsymbol{\theta}))}\hb_n\hb^T_n$,
%\begin{equation}
%\Ib(\boldsymbol{\theta})=\sum_{n=1}^N\frac{p^2_{\omega_n}(\tau_n-\hb^T_n\boldsymbol{\theta})}{F_{\omega_n}(\tau_n-\hb^T_n\boldsymbol{\theta})(1-F_{\omega_n}(\tau_n-\hb^T_n\boldsymbol{\theta}))}\hb_n\hb^T_n
%\end{equation}
where $p_{\omega_n}$ is the probability density function (pdf) of $\omega_n$. Therefore, the noncentrality parameter $\lambda_{Q}$ can be computed as
\begin{equation}
\label{eq: lambda}
\lambda_{Q}=\sum_{n=1}^N\frac{p^2_{\omega_n}(\tau_n)}{F_{\omega_n}(\tau_n)(1-F_{\omega_n}(\tau_n))}\boldsymbol{\theta}^T\hb_n\hb^T_n\boldsymbol{\theta},
\end{equation}
where $\boldsymbol{\theta}^T\hb_n\hb^T_n\boldsymbol{\theta}\geq0$. The noncentrality parameter is a function of thresholds $\tau_n$. Following \cite{Fang13} and \cite{Kay98}, a larger noncentrality parameter results in better detection performance. Hence, the optimal thresholds are those that maximize $g(\tau_n)\triangleq\frac{p^2_{\omega_n}(\tau_n)}{F_{\omega_n}(\tau_n)(1-F_{\omega_n}(\tau_n))}$. Since $\omega_n$ is a Gaussian random variable, the function $g(\tau_n)$ is a unimodal, positive and symmetric function attaining its maximum when $\tau_n=0$ \cite{Fang13}. Substituting the optimal thresholds $\tau^{*}_n=0$ back into (\ref{eq: lambda}), the largest noncentrality parameter is $\lambda_{Q}=\frac{2}{\pi}\sum_{n=1}^N\frac{(\hb^T_n\boldsymbol{\theta})^2}{\sigma^2_n}$.
%\begin{equation}
%\lambda_{Q}=\frac{2}{\pi}\sum_{n=1}^N\frac{(\hb^T_n\boldsymbol{\theta})^2}{\sigma^2_n}.
%\end{equation}
Also, with the optimal thresholds, the Fisher information matrix is given by $\Ib(\boldsymbol{\theta})=\sum_{n=1}^N\frac{p^2_{\omega_n}(-\hb^T_n\boldsymbol{\theta})}{F_{\omega_n}(-\hb^T_n\boldsymbol{\theta})(1-F_{\omega_n}(-\hb^T_n\boldsymbol{\theta}))}\hb_n\hb^T_n$.
%\begin{equation}
%\Ib(\boldsymbol{\theta})=\sum_{n=1}^N\frac{p^2_{\omega_n}(-\hb^T_n\boldsymbol{\theta})}{F_{\omega_n}(-\hb^T_n\boldsymbol{\theta})(1-F_{\omega_n}(-\hb^T_n\boldsymbol{\theta}))}\hb_n\hb^T_n
%\end{equation}
Now, the Cramer-Rao lower bound \cite{Kay02} is $var(\hat{\theta}_i)\geq[\Ib^{-1}(\boldsymbol{\theta})]_{ii}$.

Substitute (\ref{eq: pbn}) into (\ref{eq: GLRTbasic}) and note that $P(\boldsymbol b | H_0) = (\frac{1}{2})^N$, we have
\begin{equation}
\label{eq: finalGLRT}
\mathrm{GLRT}:\quad\sum_{n=1}^N b_n\alpha_n\substack{H_1 \\ > \\ < \\ H_0}\eta^{'},
\end{equation}
where $\eta^{'}=\mathrm{ln}[(\frac{1}{2})^N\eta]-\sum_{n=1}^N\mathrm{ln}(1-\beta_n)$, and $\alpha_n=\mathrm{ln}(\frac{\beta_n}{1-\beta_n})$ where $\beta_n\triangleq F_{\omega_n}(-\hb^T_n\hat{\boldsymbol{\theta}}_{ML})$. In this letter, we use the result of the BIHT algorithm $\hat{\boldsymbol{\theta}}=\hat{\boldsymbol{\theta}}_{\mathrm{BIHT}}$ instead of MLE. The GLRT detector is a weighted sum of the binary data. Since $F_{\omega_n}$ is a non-decreasing function, nodes with larger $\boldsymbol \hb_n^T \hat{\boldsymbol \theta}$ will have more influence in the overall decision making function.

To illustrate the importance of these optimal weightings, we suggest a simple uniform GLRT detector which uses $\alpha_n=1$ for all binary data of each sensor irrespective to its measurement vector $\hb_n$.

The main difference between our approach and that of presented in \cite{Fang13} is that we consider the detection of unknown sparse vector $\boldsymbol{\theta}$, while the authors in \cite{Fang13} only consider the detection of an unknown scalar deterministic signal $\theta$. Considering the sparse vector detection extends the applicability of the model. For instance, our proposed method can be applied to the spectrum sensing application (see Section VI). Also, the simplified model of the final decision for GLRT detector has not been presented in [1], while here we have presented a simple equation for GLRT final decision in (12). This in turn facilitates the implementation procedure of GLRT detector. Finally, as we deal with the sparse vector $\boldsymbol{\theta}$, we have replaced the MLE estimate of $\boldsymbol{\theta}$ with the sparsity-driven (i.e. BIHT algorithm) estimate of $\boldsymbol{\theta}$. This is because the MLE estimate does not take into account the sparsity constraint, while BIHT does and thus it leads to more accurate results.

\section{Double-detector}
\label{sec: doubleGLRT}
In this section, a threshold detection is done in each sensor based on $\hat{\boldsymbol{\theta}}$ and the binary result is sent to the FC. Then, in the fusion center, a GLRT detector is applied to the binary data received, hence the name double-detector. The hypotheses and the binary result of internal detector are given by
%\begin{equation}
%\Big\{\begin{array}{cc}
%                                     H_0:\quad x_n=\omega_n &  \\
%                                     H_1:\quad x_n=A_n+\omega_n &
%                                    \end{array}
%\end{equation}
\begin{align*}
\mathrm{H_0}:\quad x_n&=\omega_n,\\
\mathrm{H_1}:\quad x_n&=A_n+\omega_n,\quad n=1,2,...,N,
\end{align*}

\begin{equation}
c_n= \Big\{\begin{array}{cc}
                                     1\quad x_n\geq\tau_n &  \\
                                     0\quad x_n<\tau_n &
                                    \end{array},
\end{equation}
where $A_n=\hb_n^T\boldsymbol{\theta}\approx\hb_n^T\hat{\boldsymbol{\theta}}$, $\hat{\boldsymbol{\theta}}$ is the estimate of $\boldsymbol{\theta}$ from BIHT ($\hat{\boldsymbol{\theta}}\approx\boldsymbol{\theta}$) and the threshold $\tau_n=F^{-1}_{\omega_n}(P_{fa})$ where $P_{fa}$ is the false alarm probability of the internal threshold detector. Set all the internal threshold detectors to have the same false alarm probability $P_{fa}$. Then, the probability of detection of the detectors will vary $P_{dn}=F_{\omega_n}(\tau_n-A_n)$. At the fusion center, the binary results $c_n$ are received and then a final GLRT detector is used for detection of $H_1$:
\begin{equation}
\frac{P(\cb|\hat{\boldsymbol{\theta}},\mathrm{H}_1)}{P(\cb|\mathrm{H}_0)}\substack{H_1 \\ > \\ < \\ H_0}\gamma,
\end{equation}
where $\cb=[c_1,c_2,...,c_N]^T$, and $\gamma$ is a threshold determined by the final false alarm probability. As the estimate of $\boldsymbol{\theta}$ from BIHT is accurate, we have $\hat{\boldsymbol{\theta}}\approx\boldsymbol{\theta}$. Therefore, we have $P(\cb|\hat{\boldsymbol{\theta}},\mathrm{H}_1)\approx P(\cb|\boldsymbol{\theta},\mathrm{H}_1)=\prod_{n=1}^N p^{c_n}_{dn}(1-p_{dn})^{1-c_n}$ and $P(\cb|\mathrm{H}_0)=\prod_{n=1}^N p^{c_n}_{fa}(1-p_{fa})^{1-c_n}$. After straightforward calculations, the final GLRT decision is:
\begin{equation}
\sum_{n=1}^N c_n\rho_n\substack{H_1 \\ > \\ < \\ H_0}\gamma^{'},
\end{equation}
where $\gamma^{'}\triangleq \mathrm{ln}(\frac{\gamma(1-P_{fa})^N}{\prod_{n=1}^N(1-p_{dn})})$ is a threshold that controls the final false alarm probability, and $\rho_n\triangleq \mathrm{ln}(\frac{p_{dn}}{1-p_{dn}}\frac{1-p_{fa}}{p_{fa}})$ is a weighting coefficient. We see that the larger the probability of detection of internal detector, the higher weight is devoted to that detector in the final GLRT detector. 

Figure 1 shows the block digram of the two proposed detectors in this letter. Figure 1(a) represents the block diagram of the proposed sign-GLRT in Section III, where it utilizes the sign of the measurements at local sensors to quantize the data into binary data $\{b_n\}_{n=1}^N$. Then, through an ideal binary channel, the binary data is transmitted to FC. Upon receiving the binary data $\{b_n\}_{n=1}^N$, by using the BIHT estimate of the sparse vector $\boldsymbol{\theta}$ and calculating the relevant coefficients (i.e. $\{\alpha_n\}_{n=1}^N$), FC uses the GLRT detector to make final decision between the absence or the presence of sparse vector $\boldsymbol{\theta}$. Unlike the sign-GLRT detector, the proposed double-detector, presented in Fig. 1(b), performs an internal threshold detection at each local sensor (based on $\hat{\boldsymbol{\theta}}$). The binary result is then sent to FC, where a final GLRT detector is used for final decision. Hence, unlike sign-GLRT, the ML or BIHT estimate of $\boldsymbol{\theta}$ is not required at FC for final decision. Although the complexity of estimation slightly increases in the local sensors, it outperforms the sign-GLRT detector presented in Section III (see Section VI).

\section{Clairvoyant and Oracle detectors}
\label{sec: clair}
It is useful to compare the performance of the sign-GLRT (GLRT applied on the sign of the data) detector and double-detector with some impractical ideal detectors. The first is the clairvoyant detector which is a GLRT detector that have full access to the sensors' original observations \cite{Fang13}. The second is the detector that have access to the sparse vector $\boldsymbol{\theta}$. We denote this detector as oracle detector. The clairvoyant detector is:
\begin{equation}
\frac{P(\xb|\hat{\boldsymbol{\theta}},\mathrm{H}_1)}{P(\xb|\mathrm{H}_0)}\substack{H_1 \\ > \\ < \\ H_0}\delta,
\end{equation}
where $\xb=[x_1,x_2,...,x_N]^T$ is the measurement vector, $\delta$ is a threshold that determines the false alarm probability and $\hat{\boldsymbol{\theta}}=\hat{\boldsymbol{\theta}}_{ML}$ is the MLE. Fortunately, in this case, there is a closed form formula for MLE which is \cite{Kay02}:
\begin{equation}
\hat{\boldsymbol{\theta}}_{ML}=(\Hb^T\Cb^{-1}\Hb)^{-1}\Hb^T\Cb^{-1}\xb,
\end{equation}
where $\Hb$ is a $N\times M$ matrix whose $n$'th row is $\hb^T_n$, $\Cb=\mathrm{diag}(\sigma^2_n)$ is a diagonal matrix whose diagonal elements are $\sigma^2_n$, and $\xb$ follows a linear model as $\xb=\Hb\boldsymbol{\theta}+\boldsymbol{\omega}$ where $\boldsymbol{\omega}=[\omega_1,\omega_2,...,\omega_N]^T$. It can be shown that the clairvoyant detector will be:
\begin{equation}
\sum_{n=1}^N\frac{1}{\sigma^2_n}[2x_n\hb^T_n\hat{\boldsymbol{\theta}}+(\hb^T_n\hat{\boldsymbol{\theta}})^2]\substack{H_1 \\ > \\ < \\ H_0}2\mathrm{ln}(\delta).
\end{equation}
The oracle detector is similar to the GLRT detector in (\ref{eq: finalGLRT}) with $\beta_n\triangleq F_{\omega_n}(-\hb^T_n\boldsymbol{\theta})$ since we know the exact sparse vector $\boldsymbol{\theta}$.

\begin{figure}[tb]
\begin{center}
\hspace*{-0.5in}
\includegraphics[width=8cm,angle=-90]{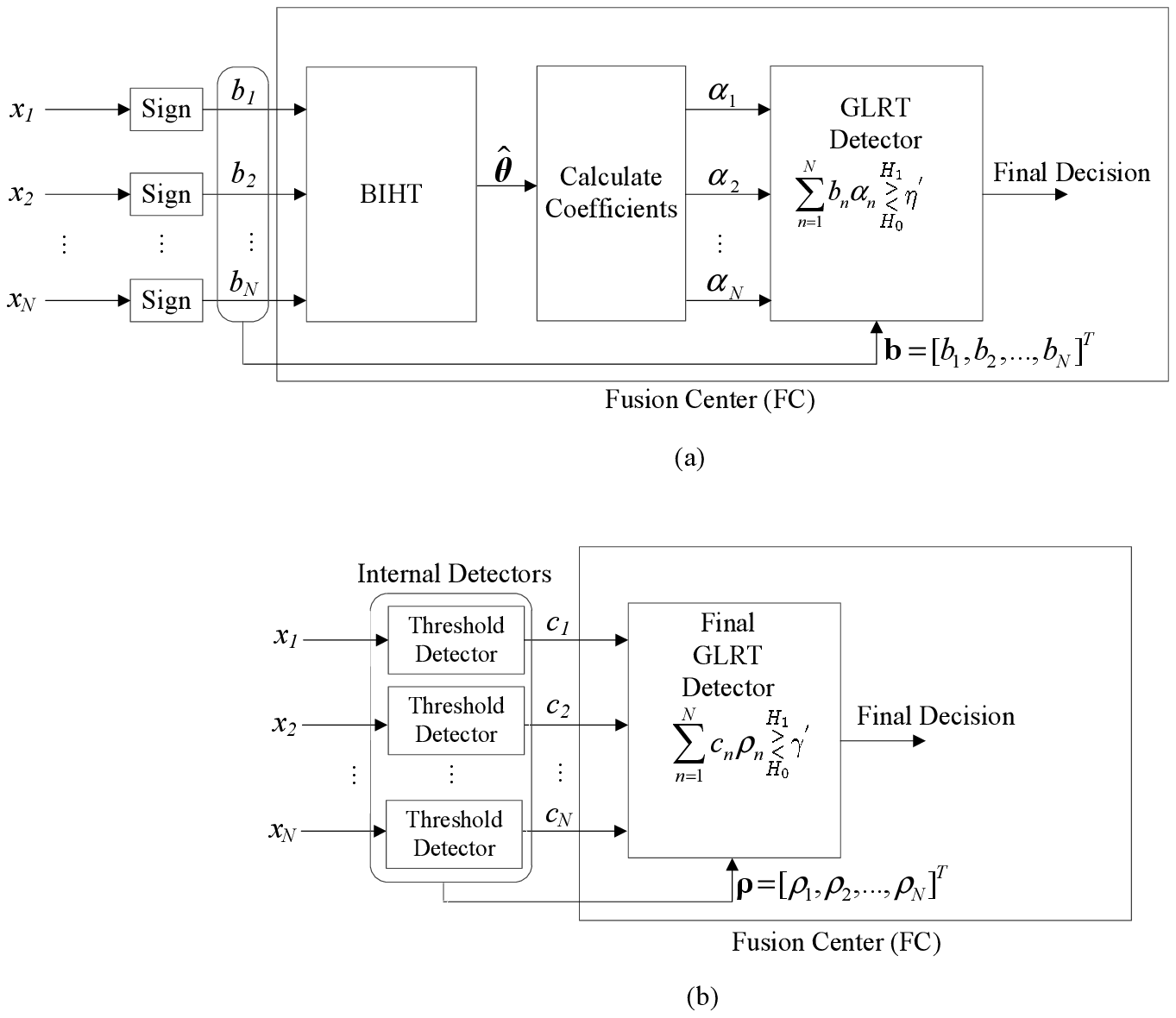}
\end{center}
\vspace{4em}
\caption{Diagrammatic representations of the proposed detectors: (a) sign-GLRT (b) double-detector. An ideal binary channel has been used for transmitting data from local sensors to FC.}
%\end{center}
\label{Fig1}
\end{figure}

\section{Simulation Results}
\label{sec: Sim}
This section presents the simulation results in support of the proposed methods. Four experiments are performed in this section. At first experiment, we examined five cases for the parameter $P_{fa}\in\{0.1,0.2,0.3,0.4,0.5\}$ of the internal detector and obtained the performance of the double-detector in these five cases in comparison to the sign-GLRT detector. The second experiment illustrates the performance of the sign-GLRT and the proposed double-detector scheme in comparison with the clairvoyant, oracle and uniform GLRT. At third experiment, the application of spectrum sensing was investigated. Finally, the computational time of the proposed detectors are compared with the existing methods. In order to have a fair comparison, we have run the sign-GLRT with two one-bit compressed sensing methods: BIHT \cite{JacqLBB13} and one-bit BCS \cite{Li15}. These are denoted as sign-GLRT (BIHT) and sign-GLRT (one-bit BCS) in the simulation results, respectively. Number of sensors are assumed to be $N=50$. The length of the sparse vector is assumed to be $M=10$ with two nonzero elements. The sparse vector is assumed to be $\boldsymbol{\theta}=[0,1,0,0,0,0,-2,0,0,0]^T$. The elements of the measurement vectors $\hb_n$ is assumed to be random Gaussian with zero mean and unit variance. The sensor's noises are assumed to have equivalent power with $\sigma^2_n=\sigma^2$. The Signal to Noise Ratio (SNR) is defined as $\mathrm{SNR}=10\log_{10}\frac{E\{(\hb^T_n\boldsymbol{\theta})^2\}}{\sigma^2}$. The experiments are repeated for $10^4$ independent runs.

\begin{figure}
\centering
\subfigure[]{
\label{figa}
\includegraphics[width=6cm]{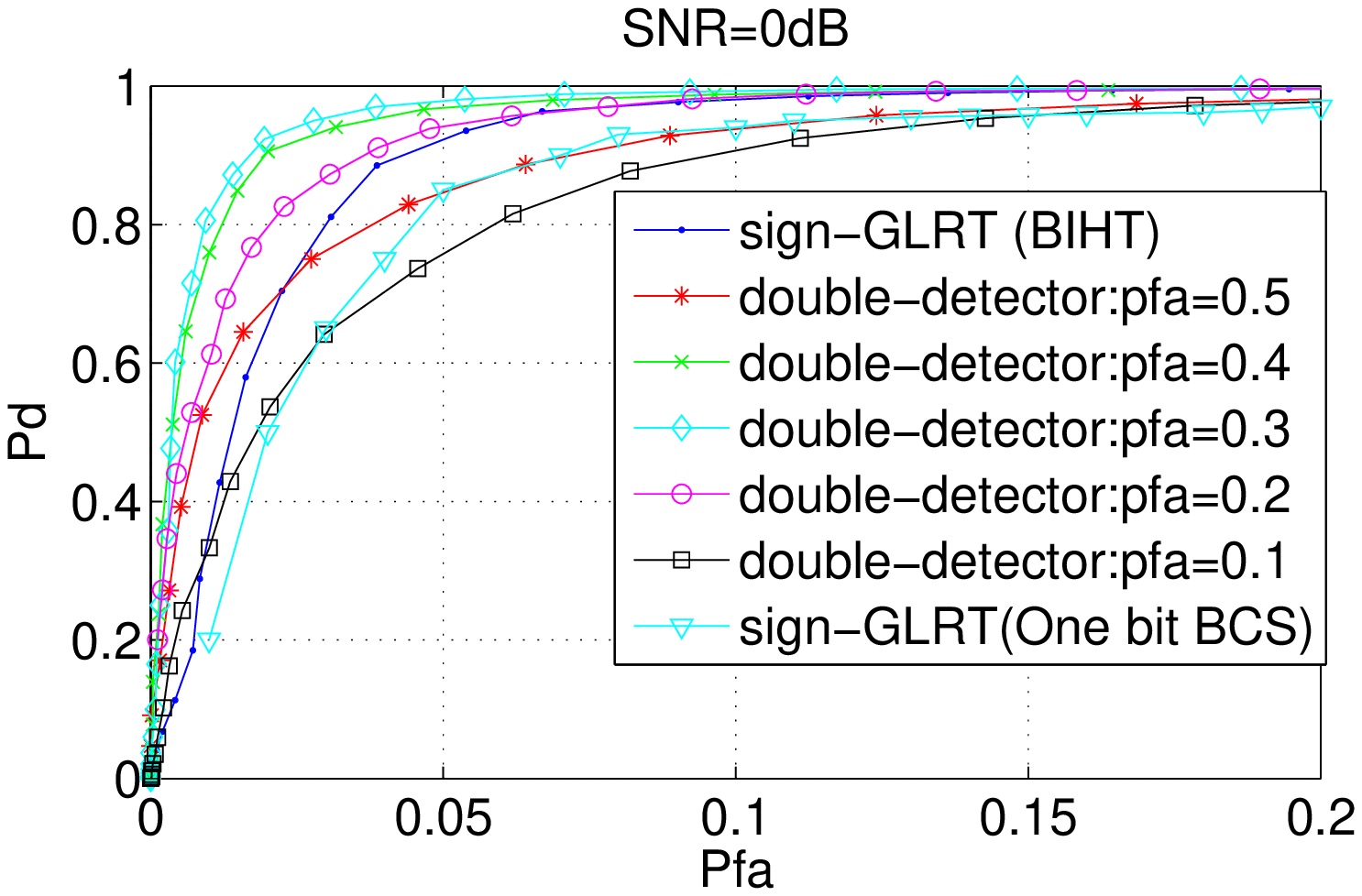}}
%\hskip 0.01\textwidth
\!
\subfigure[]{
\label{figb}
\includegraphics[width=6cm]{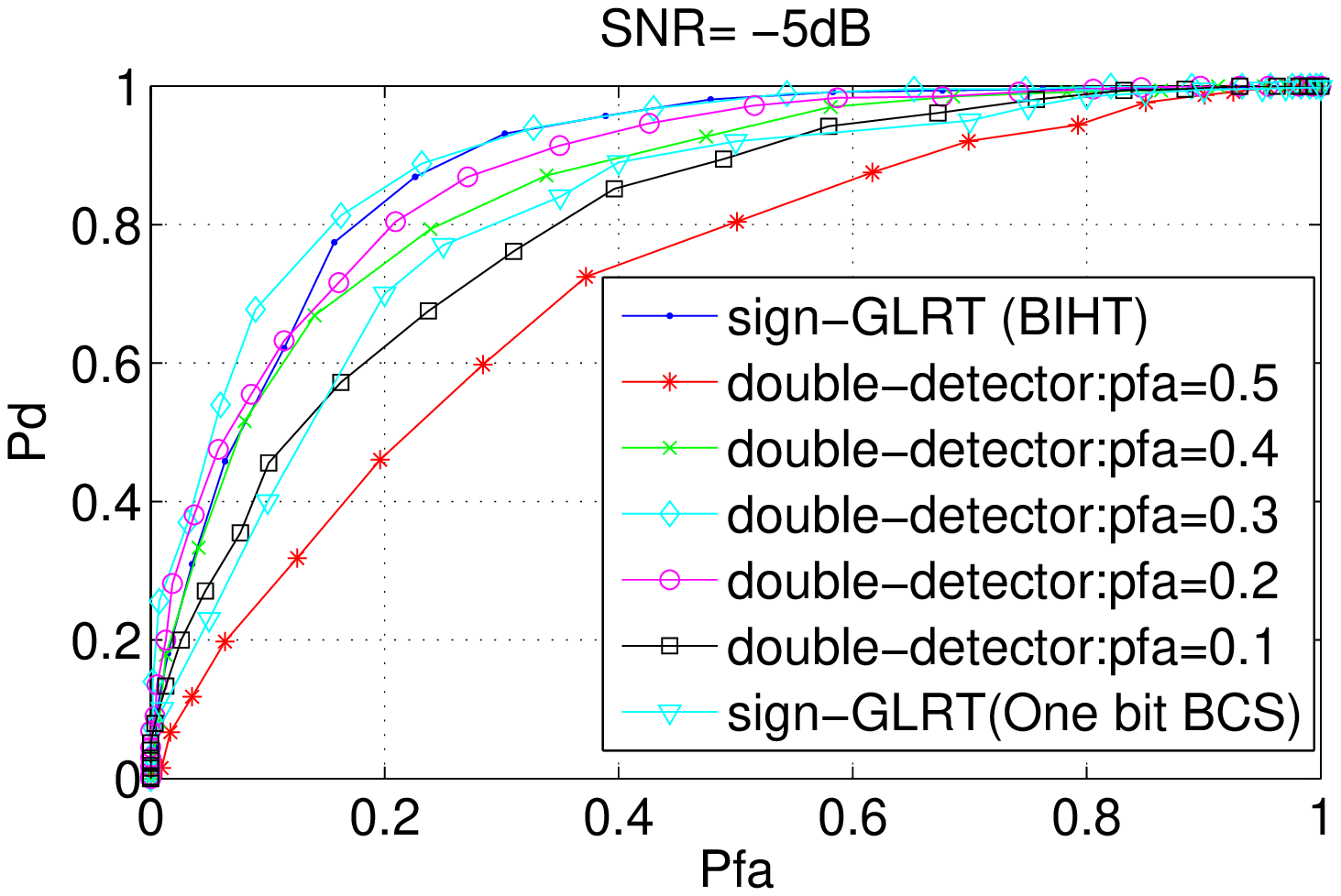}}
\caption{\footnotesize {Probability of detection versus false alarm probability for double- detector with five different values of the parameter $P_{fa}$ of the internal detector.
(a) SNR=0 dB. (b) SNR= -5dB.}}
\label{fig4}
\end{figure}

At the first experiment, performance of the double-detector was examined. Three values for the parameter $P_{fa}$ which is the false alarm probability of the internal detector were tried. Figure. \ref{fig4} shows the detection probability versus false alarm probability of double-detector in five cases which are $P_{fa}\in\{0.1,0.2,0.3,0.4,0.5\}$ in comparison to the sign-GLRT (BIHT), sign-GLRT (one-bit BCS) and in two cases for SNR. It shows that the best performance among these five cases is achieved by $P_{fa}=0.3$. Therefore, we chose this value for the next experiments.

\begin{figure}[tb]
\begin{center}
\includegraphics[width=6cm]{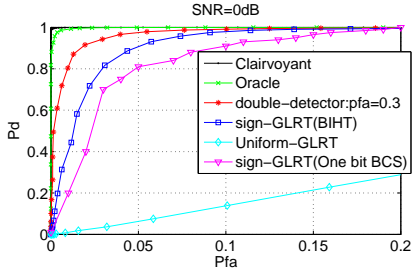}
\end{center}
\caption{Probability of detection versus false alarm probability for sign-GLRT, double-detector, uniform GLRT, clairvoyant and oracle detectors.}
%\end{center}
\label{fig2}
\end{figure}

At the second experiment, performance of the double-detector with the parameter $P_{fa}=0.3$ was compared with the sign-GLRT (BIHT), sign-GLRT (one-bit BCS), uniform GLRT, clairvoyant and oracle detectors. The detection probability versus false alarm probability of these detectors are shown in Fig.~\ref{fig2}. It shows that the double-detector outperforms both sign-GLRT (BIHT) and sign-GLRT (one-bit BCS). The clairvoyant has the best performance and the uniform GLRT which uses the same weights for all sensors has the worst result.

\begin{figure}
\centering
\subfigure[]{
\label{figa}
\includegraphics[width=5cm]{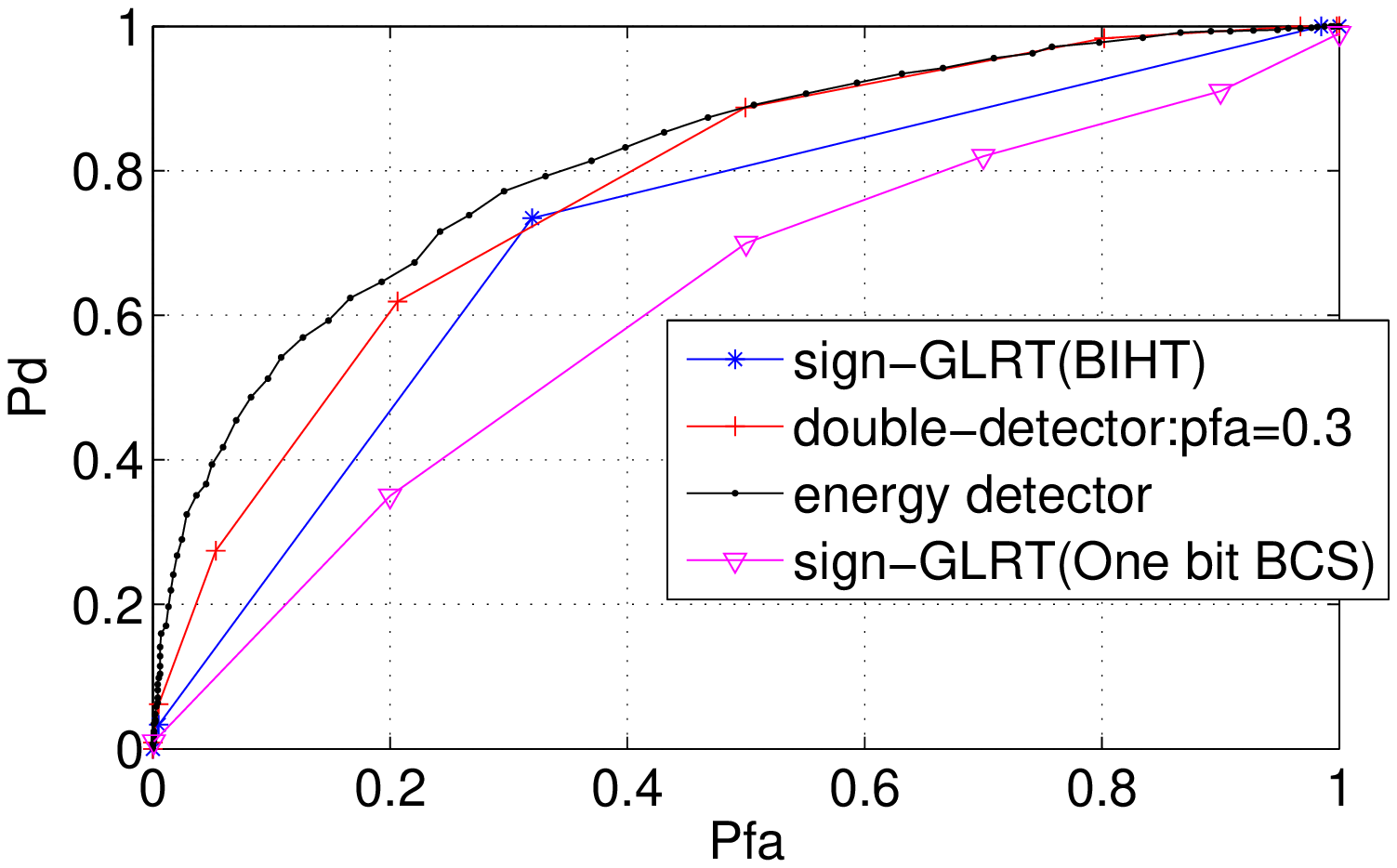}}
%\hskip 0.01\textwidth
\!
\subfigure[]{
\label{figb}
\includegraphics[width=5cm]{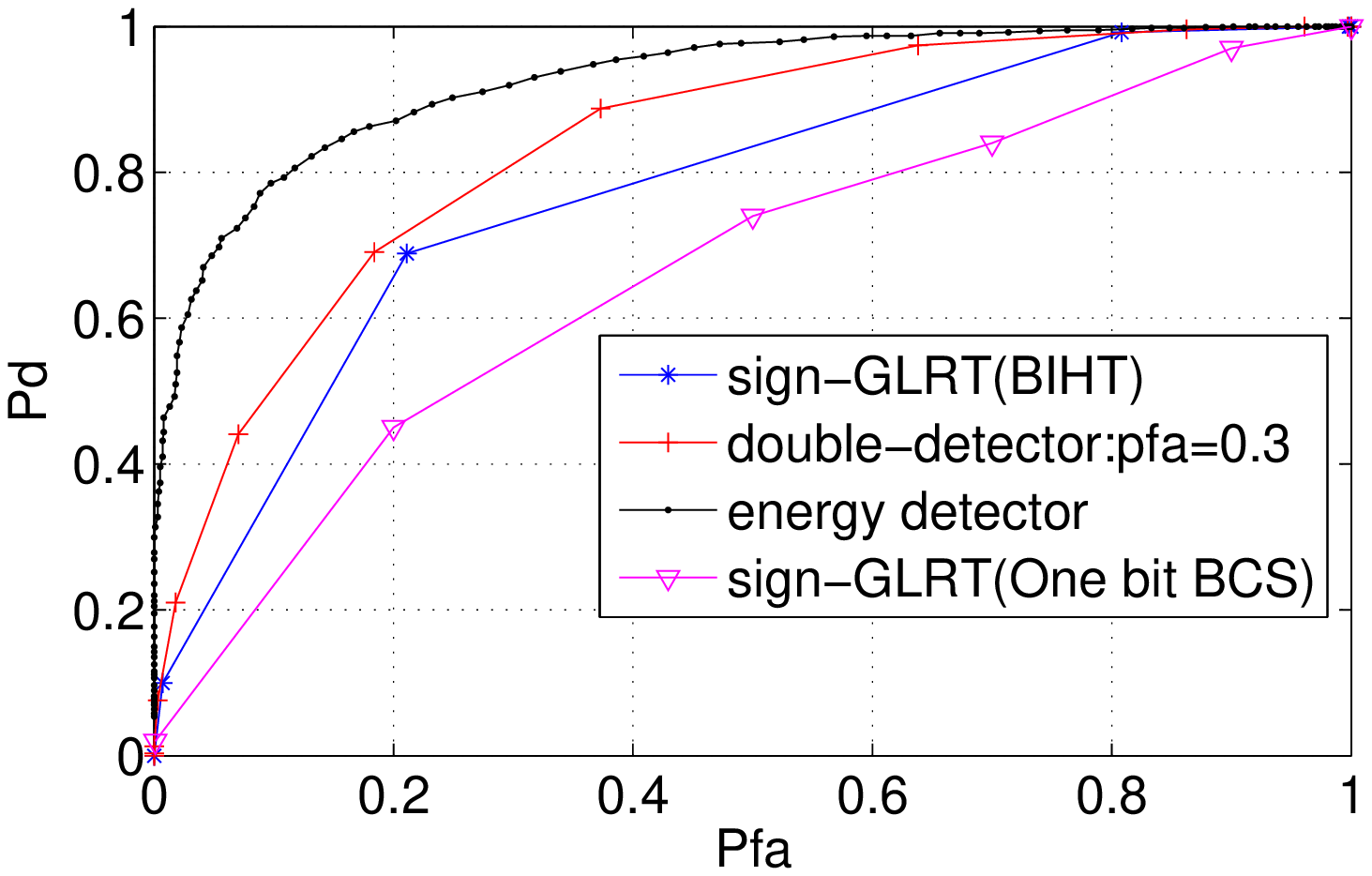}}
\caption{\footnotesize {Probability of detection versus false alarm probability at SNR=0dB in spectrum sensing for double-detector and sign-GLRT in comparison to energy detector.
(a) N=40. (b) N=100.}}
\label{fig5}
\end{figure}

At the third experiment, the application of spectrum sensing is considered. A real synthetic signal $\boldsymbol{\theta}$ with the length of $M=128$ is generated whose discrete fourier transform has only three non zero components at frequency bins indexed in $\{10,20,30\}$ with the frequency amplitude equal to $\{1.0,0.5,2\}$, respectively. Therefore, signal vector can be represented in the matrix notation as $\boldsymbol{\theta}=\Fb^{-1}\boldsymbol{\psi}$, where $\Fb^{-1}$ is the inverse Fourier matrix and $\boldsymbol{\psi}$ is the discrete Fourier transform and is sparse with only three non zero elements of 128 elements. Hence, the sensor measurements are $x_n=\hb^T_n\boldsymbol{\theta}+\omega_n=\hb^T_n\Fb^{-1}\boldsymbol{\psi}+\omega_n={\hb^{'}}^T_n\boldsymbol{\psi}+\omega_n$, where ${\hb^{'}}^T_n=\hb^T_n\Fb^{-1}$. Now, the model is the same as our model and the presence of interference in the spectrum is equivalent to presence of the sparse signal. Therefore, the double-detector, sign-GLRT (BIHT), and sign-GLRT (one-bit BCS) detectors can be used for spectrum sensing. The well-known spectrum sensing detector is the energy detector which is applied to the original real valued data obtained from sensors. The sensors send the real valued data (not a sign) to a fusion center and fusion center detects the presence of the signal when $\sum_{n=1}^N|x_n|^2>\mathrm{Th}$ where $\mathrm{Th}$ determines the false alarm probability. The SNR is selected as $\mathrm{SNR}=0dB$ and two cases for the number of sensors are considered which are $N=40$ and $100$. The results are shown in Fig~\ref{fig5}. It shows that the detection performance of double-detector is better than both sign-GLRT (BIHT) and sign-GLRT (one-bit BCS). Also, the performance of the proposed detectors, i.e. sign-GLRT (BIHT) and double-detector, is close to that of energy detector while the structure of the wireless sensor network could be very simple and less demanding with respect to power and bandwidth.

Finally, we compare the computational time of the proposed detectors with existing methods in the setting of second experiment. The average runtime of 1000 runs of the sign-GLRT (BIHT) is 0.0046 seconds while the runtime of sign-GLRT (one-bit BCS) is 0.0122 seconds. The double-detector which does not use any one-bit compressed sensing algorithm at FC has a runtime equal to 0.0036 seconds. The proposed double-detector is faster than all the other detectors.

%At the third experiment, three noisy cases were examined. The variance of measurement noises $\sigma_n$ is selected as $\sigma_n=.1$, $\sigma_n=1$ and $\sigma_n=2$. The ROC curves for the sign-GLRT and the double-GLRT are compared with each other in Fig~\ref{fig3} and in these three cases.

%\begin{figure}[tb]
%\begin{center}
%\includegraphics[width=8cm]{exp_pfa.jpg}
%\end{center}
%\caption{Probability of detection versus false alarm probability for double GLRT with three different values of the parameter $P_{fa}$ of the internal GLRT.}
%%\end{center}
%\label{fig1}
%\end{figure}

%\begin{figure}[tb]
%\begin{center}
%\includegraphics[width=6.8cm]{exp2_final.jpg}
%\end{center}
%\caption{NMSE versus number of sign measurements.}
%%\end{center}
%\label{fig3}
%\end{figure}

%\begin{figure}[tb]
%\begin{center}
%\includegraphics[width=8cm]{exp3_final.jpg}
%\end{center}
%\caption{NMSE versus Signal to Noise Ratio.}
%\end{center}
%\label{fig4}
%\end{figure}

\section{Conclusion}
\label{sec: con}
The problem of vector sparse signal detection from one bit compressed sensing measurements was addressed. It is a generalization of the problem of a scalar signal detection from one bit measurements. The GLRT detector and the optimal quantizer thresholds were derived. The GLRT detector was simplified to an intuitive criterion for detection of the sparse signals. A double-detector is also proposed to improve the sign-GLRT detector. In this new scheme, the decisions of the internal detectors are transmitted to the fusion center and a final GLRT detector detects the sparse signal. Moreover, a clairvoyant detector of the problem and an oracle detector were introduced. Simulation results show that by careful designing of the internal detector in the double-detector scheme, a better performance than sign-GLRT detector can be obtained. Also, application of the suggested detector in spectrum sensing shows slightly worse result than energy detector while structure of the wireless sensor network could be very simple and low power (one bit A/D in instead of multi bit A/D's).

\ifCLASSOPTIONcaptionsoff
  \newpage
\fi

\end{document}